\documentclass[preprint]{aastex}
\usepackage{xcolor}

\shorttitle{A New Generation of Los Alamos Opacity Tables}
\shortauthors{Colgan et al.}

\begin{document}


\title{A New Generation of Los Alamos Opacity Tables}


\author{J.~Colgan, D.~P.~Kilcrease, N.~H.~Magee, M.~E.~Sherrill, J.~Abdallah, Jr., P.~Hakel, C.~J.~Fontes, J.~A.~Guzik, K.~A.~Mussack}
\affil{Los Alamos National Laboratory, Los Alamos, NM 87545}
%
%

\begin{abstract}

We present a new, publicly available, set of Los Alamos OPLIB opacity tables for the elements hydrogen through zinc. 
Our tables are computed using the Los Alamos ATOMIC opacity and plasma modeling code, and make use
of atomic structure calculations that use fine-structure detail for all the elements considered. Our equation-of-state
(EOS) model, known as ChemEOS, is based on the minimization of free energy in a chemical picture and appears
to be a reasonable and robust approach to determining atomic state populations over a wide range of temperatures
and densities. In this paper we discuss in detail the calculations that we have performed for the 30 elements considered,
and present some comparisons of our monochromatic opacities with measurements and other opacity codes. We also
use our new opacity tables in solar modeling calculations and compare and contrast such modeling with previous work.

\end{abstract}

\keywords{opacities, atomic data, solar mixture}

\section{Introduction}

The radiative opacity is a fundamental property of a material that determines the amount of radiation absorbed and scattered \citep{huebner}.
In general, the opacity is dependent on the radiation temperature, material temperature and density of the material as well as the wavelength of the incoming radiation. 
The knowledge of material opacities are crucial in determining the transport of radiation through a material and therefore both quantities play a major
role in stellar modeling including stellar evolution, pulsation, and in determining large-scale stellar quantities such as elemental abundance,
temperature profiles, etc. Thermodynamic equilibrium is reached when the material and the radiation are at the same temperature and 
populations are in steady-state, a scenario often encountered in stars, and this
allows the ready determination of material emissivity from the opacity via Kirchhoff's law. Stellar models have thus relied on tables of opacities
computed in local thermodynamic equilibrium (LTE) for a range of elements present in the system of interest. Long-term intensive efforts
to produce accurate and comprehensive opacity tables have been underway for many years, with notable efforts being the Opacity
Project (OP) \citep{op0,op1,op2}, the Lawrence Livermore National Laboratory OPAL opacity tables \citep{opal1,opal2}, and more recently
opacity tables produced using the OPAS code \citep{opas}. {\color{red} The OPAS code has very recently been used to compute opacities
for solar mixtures \citep{mondet,lepennec2015}, and improved agreement with helioseismic observations was reported. The SCO-RCG code \citep{scorcg} also appears to be a powerful method
with which to compute opacities.} At Los Alamos National Laboratory, tables of opacities have been computed using
the LEDCOP code \citep{ledcop} in the 1990s and these OPLIB tables have successfully been used in solar modeling \citep{guzik1,guzik2}. 

In recent years, it has become apparent that more refined opacity calculations could be useful in stellar modeling. In particular, the
fairly recent discovery that the revision in the solar elemental abundances \citep{asplund05} has destroyed the previously good agreement
that existed between standard solar model predictions made using older solar elemental abundances \citep{gn93} and helioseismic observations \citep{bahcall05} has led to renewed scrutiny of
the opacities used in such solar models. It was quickly noted \citep{serenelli09} that an increase in opacity of some of the major solar elements of between 5--20\% in the solar radiative zone would restore the agreement between helioseismology and the solar models, although the two main sets of tables used in solar modeling (OP and OPAL) are in reasonably close agreement in this zone. Although even more recent studies
have slightly revised the new solar abundances \citep{asplund09}, the `solar abundance problem' is still not resolved \citep{guzik10}.
Opacities have also been postulated as the source of discrepancies between observations of pulsations of $\beta$-Cepheids and modeling,
and in particular significant differences have been found between models when either OP or OPAL opacity tables are used \citep{dd09,dd10,dd13,cugier}. More recent work by \citet{walczak15}, which uses the new opacity tables described in this paper, produces
improved agreement with observed pulsations for B stars.

These considerations, coupled with larger computational resources and more robust physical models, have led to a new opacity effort
at Los Alamos National Laboratory (LANL) using the ATOMIC code \citep{magee,hakel}. The aim of this effort is to supplant the existing
LANL OPLIB opacity tables previously computed  using LEDCOP with a new generation of opacity tables computed using ATOMIC. This effort has
been completed for all elements from hydrogen through zinc and our tables are available online\footnote{http://aphysics2.lanl.gov/opacity/lanl}. 
The purpose of this
paper is to describe in detail calculations that have been performed in generating these new opacity tables and to compare and contrast
our new opacities with available measurements and previous theoretical work. Several previous publications \citep{colgan1,colgan2,colgan3}
have described a few aspects of our ATOMIC calculations, including comparisons \citep{colgan3} of the monochromatic opacity against several sets of opacity measurements made in the 1990s \citep{foster,springer,perry,winhart}. 
In this paper we also discuss comparisons of our opacity calculations 
against the opacity measurements  made using the Sandia National Laboratory Z-pinch machine \citep{bailey07,bailey15}.  Finally,
we also present new solar mixture opacity calculations and discuss the implications of the new opacities in solar modeling.

\section{Theoretical Methods}

Our opacity calculations were made using the ATOMIC code. ATOMIC is a multi-purpose plasma modeling code \citep{magee,hakel,fontes15} that can be
run in LTE or non-LTE mode. A major strength of ATOMIC is the ability to easily run at various levels of refinement \citep{fontes16}. For example, depending
on the atomic datasets available, one can run with atomic data generated in the configuration-average approximation, or in
fine-structure detail. The atomic data used in this work were generated with the semi-relativistic capability
in the Los Alamos suite of atomic physics codes. An overview of this capability
has recently been provided by \citet{fontes15}.

The overall aim of our new opacity calculations is to compute a set of monochromatic opacities and Rosseland mean opacities for the 
elements hydrogen through zinc and for a wide range of temperatures and densities. We provide opacities over a temperature ($T$)  range of 
0.5~eV up to 100~keV, and for mass densities that span at least 12 orders of magnitude, starting at mass densities of around 
$10^{-8}$ g/cm$^3$ or lower (depending on the temperature and element under consideration).
We also provide Planck mean opacities, although
our focus is primarily on the Rosseland mean opacity since it is usually the quantity of main interest in most astrophysical applications.
We define the Rosseland mean opacity as \citep{cg04}
\begin{eqnarray}
\frac{1}{\kappa_{\rm ROSS}} = \frac{\int_0^{\infty} \frac{1}{\kappa_{\nu}} n_{\nu}^3 \frac{\partial B_{\nu}}{\partial T} d\nu}{\int_0^{\infty} \frac{\partial B_{\nu}}{\partial T} d\nu} \ ,
\label{rossop}
\end{eqnarray}
where $\nu$ is the photon frequency, $B_{\nu}$ is the Planck function \citep{huebner}, $n_{\nu}$ is the frequency-dependent refractive index
defined by \citet{armstrong}, and $\kappa_{\nu}$ is the frequency-dependent opacity. Our opacities are given in cm$^2$/gram.
  $\kappa_{\nu}$ is composed of various contributions
that can be summarized as
\begin{eqnarray}
\kappa_{\nu} = \kappa_{\rm BB} + \kappa_{\rm BF} + \kappa_{\rm FF} + \kappa_{\rm SCAT}  \ ,
\end{eqnarray}
i.e. the sum of bound-bound (BB), bound-free (BF), free-free (FF), and scattering (SCAT) contributions. The first three of these contributions
include a factor due to stimulated emission.
In the following subsections we discuss in detail the various calculations of each of these contributions as well as several related issues.

\subsection{Choice of configuration model for opacity calculations}

For a complete opacity calculation it is crucial to include sufficient numbers of states (configurations, levels, etc) so that the
calculation is converged with respect to contributions to the total absorption. In the ATOMIC calculations presented here, we used
a similar set of configurations that were used in the older LEDCOP calculations. These configurations are then split
into fine-structure levels by one of several methods as discussed in the following subsections. The new sets of configurations were initially
chosen based on 
considerations of what configurations would be likely to retain significant population for the (large) density and temperature range
over which the tables run. A general rule of thumb was that all configurations that had an energy within about two times the ionization
energy of the ground configuration were included. From this list of configurations, we also included configurations that had one-electron
excitations from the valence sub-shell up to the $nl \equiv 10m$ subshell. Finally, in order to obtain reasonably complete bound-bound
contributions over wide photon
energy ranges, we also included one-electron dipole-allowed promotions from {\it all} subshells of this list of configurations. This final list of 
configurations was then used in the calculations of the EOS and opacity.
Some specific examples follow.

For H-like ions, the list of configurations is straightforward, and includes all configurations from $1s$ through $10m$.
For a case with many more electrons, such as neutral Sc (ground configuration: [Ar]$3d^1 4s^2$, where [Ar] means the electron configuration corresponding to the ground state of neutral Ar) we choose configurations of the type (where now the Ar core is omitted in this listing):
$3d^1 4s^2$, $3d^1 4s^1 nl$,
$3d^1 4p^2$, $3d^1 4p^1 nl$,
$3d^1 4d^2$, $3d^1 4d^1 nl$,
$3d^1 4f^2$, $3d^1 4f^1 nl$,
$3d^3$, $3d^2 nl$,
$4s^2 nl$,
$4s^1 4p^2$, $4s^1 4p^1 nl$,
$4p^3$, $4p^2 nl$.
Again, $nl$ extends up to $10m$. Contributions from levels with $n>10$ are included in an approximate manner (see following sections).
This choice of configurations encompasses 384 configurations, and these are then used to promote an
electron from each sub-shell to any (open) sub-shell that is available via a dipole promotion (i.e. where the orbital angular momentum of
the jumping electron changes by $\pm 1$). For neutral Sc, this choice leads to a list of 21,144 configurations. Frequently, we find lists
of configurations that are significantly larger than this, especially for ions with (near) half-filled shells in their ground configuration. For example,
the total number of configurations employed for the Fe opacity table was more than 600,000 (for all ion stages of Fe). 

We note that this prescription for choosing configurations does not guarantee that all possible configurations that may significantly contribute to an
opacity are included. However, we believe that our approach is sufficiently inclusive that it is likely that more configurations are retained than needed
for any given temperature and density, although this is difficult to explicitly demonstrate for every temperature/density point considered without even larger calculations. In Section 3 we do discuss
the effect of increasing the number of configurations for one set of opacity calculations for Fe.
We do note that the relative ease of choosing sets of configurations, and the guarantee of consistency between and within
datasets, make this approach to choosing configurations  viable for the generation of large-scale opacity tables.

\subsection{Bound-bound opacity contribution}

In the opacity calculations presented here, we make use of atomic data generated in a variety of ways. 
All atomic structure calculations were performed using the Los Alamos CATS code \citep{cats}, a modern version of Cowan's codes
\citep{cowan}. The calculations were carried out using the semi-relativistic Hartree-Fock, or HFR, option. 
For the Li-like, He-like and H-like
ions of all elements discussed here, we use atomic data generated at a fine-structure level of detail including full configuration-interaction
between all configurations. For ion stages with more electrons than Li-like, inclusion of full configuration-interaction proved too
computationally expensive. Instead, we employed a `single-configuration' approximation, in which we include mixing among the pure $LSJ$
basis states
that arise from a given configuration (also known as intermediate-coupling; see \citet{fontes15} for details). Oscillator strengths are computed from wavefunctions
containing the same limited amount of mixing. These calculations are accomplished in CATS by looping over
transitions occurring between pairs of dipole-allowed configurations, which automatically exclude matrix elements between pure $LSJ$
basis states that would arise from configuration-interaction.
This option, by omitting the mixing between different configurations, greatly
reduces the run time for large structure calculations, while retaining the total number of $LSJ$ levels in the
calculation. To improve the accuracy of $\Delta n=0$ transitions for $L$-shell ions, we perform a small configuration-interaction calculation
for all possible $1s^2 [2]^w$ configurations, where $[2]^w$ represents all possible permutations of $w$ electrons in the $n=2$ shell. The resulting energies replace the (less accurate) energies
for those levels computed within the single-configuration calculation, resulting in improved $L$-shell line positions in the frequency-dependent opacity.
The above procedure results in a comprehensive set of level-resolved structure and oscillator strength data and was implemented for
the Be-like through neutral stages of the elements Be through Si.

For elements beyond Si (i.e. P through Zn) we used a mixed-UTA (MUTA) approach \citep{ma} for all ions from Be-like through the neutral stage.
The MUTA method  retains all of the strongest fine-structure lines in a given transition array,
which allows an accurate spectral description to be generated from a set of configuration-average populations. This approach allows
single-configuration fine-structure detail to be included in a
relatively inexpensive computational manner. The generation of atomic data for a full level-resolved calculation 
is too computationally intensive at present. In the calculations
presented here, we retained all fine-structure lines for transition arrays that contained less than $10^5$ lines within the array. 
We have found that, for almost
all conditions of interest, this parameter choice allows essentially all lines of importance to be explicitly included in the bound-bound opacity
contribution.
A histogram approach that was introduced by \citet{abd07} to speed up computation of bound-bound contributions to spectra was also 
modified to work with the MUTA approach and was found to significantly speed up large-scale calculations with essentially no loss in 
accuracy. The histogram approach was used for the computation of opacity tables for all elements apart from H, which 
did not have significant run-times and so did not require this speed up option.
The MUTA approach has been compared to previous Fe
transmission measurements, where excellent agreement was obtained \citep{bailey07} for a temperature of 160~eV and an electron density of a few times $10^{21}$ cm$^{-3}$. Testing shows that our use of the MUTA approximation appears to be accurate for temperatures
above around 10~eV. Below this temperature, the assumption of statistical splitting of the population of a configuration into its constituent
levels may not be as accurate. This is, in part, due to the increasing importance of the Boltzmann factor $e^{-\Delta E/kT}$ at lower 
temperatures, which appears in the expression for calculating a given population. Using the configuration-average transition energy, 
$\Delta E_{CA}$, may not be a good approximation for the associated
fine-structure transition energies for some types of transitions.

\subsection{Bound-free opacity contribution}

The bound-free contribution to the opacities presented here are computed using configuration-average distorted-wave photoionization cross sections
calculated using the GIPPER ionization code \citep{gipper}. We include photoionization contributions between all possible configurations
in neighboring ion stages. We also include bound-free contributions from the photodetachment of H$^-$ using the data provided by
\citet{gelt62}.
The occupation probabilities that are discussed in Section 2.7 are used to model the merging of a Rydberg series with its corresponding
bound-free edge \citep{hub94,dappen87}. This `edge blending' approach results in smoother monochromatic opacities near a bound-free
edge by accounting for the redistribution of population from bound to continuum states via pressure ionization, while still 
conserving oscillator strength. The edge blending procedure fills the gap between the highest included $n$-value and the original edge, and
is performed down to the lowest $n$ value of the Rydberg series under consideration.

\subsection{Free-free opacity contribution}

The free-free contribution to the opacity computed by ATOMIC (also known as inverse Bremsstrahlung) is computed using
the tables provided by \citet{itoh}, results that are correct for any plasma degeneracy,
and corrected for plasma screening based on methods developed by \citet{green58,green60,armstrong}. 
This contribution is also corrected at low frequencies by incorporating effects due to multiple electron-ion collisions  \citep{iglesias10}. We also include free-free contributions from the H$^-$ \citep{gelt65}, He$^-$ \citep{somer} and C$^{-}$ ions \citep{bell}.
These latter contributions are important only at low temperatures below 1.0~eV or so. We note that the inclusion of the free-free 
contribution used in ATOMIC is very similar to that in LEDCOP, apart from our use of the multiple electron-ion collision correction at low
frequencies. The C$^-$ free-free contribution was also not included in the LEDCOP calculations. A recent study by \citet{armstrong}
used an ab-initio partial wave expansion (and more computationally intensive) approach to compute the free-free contribution. It was found that similar results were obtained compared to the use of the tables of \citet{itoh} for most conditions of interest.

\subsection{Photon scattering opacity contribution}

For photon scattering from free electrons, we use the early tables of Sampson \citep{samp} for Compton scattering, modified to include the effects of Pauli blocking and collective effects, and also include plasma non-ideality effects \citep{pb_paper} due to strong coupling and electron
degeneracy. 
The use of Compton scattering lowers the opacity at high photon energies as compared to the use of
Thomson scattering. The inclusion of Pauli blocking and collective effects
lead to a decrease in the Rosseland mean opacity at very high densities. We also note
that, when computing the number of free electrons for Compton scattering, we consider bound electrons as free when the 
photon energy is equal to or greater than the electron's binding energy. 
Finally, we also include the effects of Rayleigh scattering for the low energy side of the lowest energy transition of the ground state
of all  neutral atoms. The formulae used for this contribution for various neutral species are given in the Appendix. We note that at
low temperatures, where neutral species contributions often dominate the opacity, the Rayleigh scattering contribution can be very important.

\subsection{Line broadening}

We now discuss the line broadening packages used in ATOMIC. As is well known, 
Stark broadening of H-like and He-like  lines are important in the opacity from such ions, and this was included following the
procedure of \citet{starklee}. This procedure was modified within ATOMIC to make use of atomic data computed from our atomic structure
calculations (i.e. CATS \citep{cats}), rather than from data tables as originally proposed by \citet{starklee}.
To make our calculations
completely consistent, we included natural broadening within the Stark broadening package, and extended the temperature and density ranges
over which Stark broadening is included. We also introduced neutral resonance and neutral van der Waals broadening,
that is line broadening contributions due to the presence of other neutral
atoms within the plasma through van der Waals interactions \citep{vdw,nb} and through resonance effects \citep{ali1,ali2}, into our line broadening package.
These processes were found to give a small contribution
in low temperature regions, where neutral species dominate. It was especially relevant for He, where the large ionization
potential of the neutral atom results in a fairly wide temperature range over which neutral contributions to the opacity are important.
For lines not treated by the Stark package we use a Voigt profile incorporating Doppler, natural, and, where applicable, neutral broadening.
Electron collisional broadening is included using the approach of Armstrong \citep{arm1}. 

A detailed study was also made of line shapes far from the line center. It was found that, again particularly for He, for cold temperatures
the line-wing of the nearest bound-bound transition continues to dominate the opacity, even $10^6$ half-widths away from the line center.
The question then arises as to what is the correct form of the line shape in such a region. Previous studies by \citet{op0} and
\citet{griem_book} reached differing conclusions as to the form of this line shape. After some further consideration and based on the
arguments of \citet{heitler}, the choice was made that in the far red line wing of all absorption lines, the $\omega^4$ behavior
used by \citet{op0} in the OP data was the most
suitable, where $\omega$ is the photon angular frequency.
This decision was based on consideration of the scattering and absorption processes within a QED framework.
We note that this line shape feature can make a very large difference to the opacity  in certain regions. For example, if the
$\omega^4$ line shape is not used, the Rosseland mean opacity for He at low ($\sim 1$ eV) temperatures, may increase by more
than three orders of magnitude, since the far line wing of the nearest bound-bound transition is (essentially) the only contributor to
the opacity at these conditions. At larger temperatures, and for other elements with smaller ionization potentials, this effect is
much smaller. 

\subsection{Equation-of-state used in opacity calculations}

The equation-of-state (EOS) model used in ATOMIC is known as ChemEOS \citep{chemeos1,chemeos2}, which we summarize here. 
This approach is 
based on the minimization of the Helmholtz free energy in the chemical picture. Adopting this approach allows us to write the total
free energy as
\begin{eqnarray}
F = F_1 + F_2 + F_3 + F_4 + F_5 \ ,
\end{eqnarray}
where in this equation $F_1$ represents the ideal gas of atoms and ions, $F_2$ represents the contribution associated with the internal
energy of the atoms and ions, and $F_3$ represents the ideal Fermi electron gas free energy term. The internal energy $F_2$ term, given by
\begin{eqnarray}
F_2 = \sum_{s\neq e} N_s \left( E_{s1} - kT \; {\rm ln} \tilde{Z_s} \right) \ ,
\end{eqnarray}
depends on the converged partition function defined as
\begin{eqnarray}
\tilde{Z_s} = \sum_j w_{sj} \; {\rm exp} \left( -\frac{E_{sj}-E_{s1}}{kT} \right) \ ,
\end{eqnarray}
where in these equations $N_s$ is the number of particles of species $s$ (not including electrons), 
$kT$ is the temperature, and $E_{sj}$ is the $j$th-state energy. The convergence of the partition function is ensured
through the occupation probabilities $w_{sj}$ that smoothly truncate the summation by progressively reducing the effective
statistical weights of the excited states due to their perturbation by plasma effects. 
They are given by $w_{sj} = w_{sj}^{HS} Q_s(\beta_{sj})$, where $w_{sj}^{HS}$ is a first-order
hard-sphere contribution based on the size of the bound state \citep{hm88}, $Q_s$ is the cumulative microfield distribution function with
$\beta_{sj}$ the critical microfield \citep{pcg02}. For H and He we use a screened microfield distribution \citep{pcg02}, but for all
other elements we use an unscreened microfield distribution due to the prohibitive cost of the screened distribution computation for systems
with large numbers of states. 

The hard-sphere occupation probability term $w_{sj}^{HS}$ is included for all atoms and ions, and results in the elimination of the
(unphysical) atomic state populations at high densities.
We also note that our partition function is extended from the $n=10$ principal quantum number value explicitly included in our 
list of configurations up to $n=100$ via analytic quantum defect terms. This extension was also performed in the previous LEDCOP calculations.

The $F_4$ term in Eq.~(1) contains the Coulomb contributions \citep{cp98,pc00} to the free energy and is broken up into three
contributions: ion-ion, ion-electron, and electron-electron contributions. The ion-electron term within $F_4$ was recently modified in
\citet{chemeos2} to ensure that the effect of electron-ion binding is more consistently taken into account. The 
Coulomb interaction terms  of \citet{cp98,pc00} include electron degeneracy to all orders and our modification
also takes this electron degeneracy into account. The final term in Eq.~(1), $F_5$, accounts
for the finite size of the atom or ion through 
an excluded volume effect using an all-order hard sphere packing term. Detailed discussion of all these contributions can be found in \citet{chemeos1}. 

\subsection{The Los Alamos National Laboratory Opacity Website}

For over a decade, the Los Alamos OPLIB opacity tables have been accessible via a website: \\
{\it http://aphysics2.lanl.gov/opacity/lanl}. This webpage has
been updated recently to include access to the new OPLIB opacity tables that have been computed using the ATOMIC code. On the webpage,
the user may request opacities for single elements or mixtures of elements (any arbitrary mixture of the elements hydrogen through zinc
may be specified).  The user may obtain monochromatic opacities (these include opacities on a temperature-scaled $u=h\nu/kT$
grid of 14,900 photon energies that are
chosen to encompass a large photon energy range, and to provide a sufficient density of points in the region where the Rosseland
weighting function is peaked), multigroup opacities, or Rosseland mean and Planck mean opacities. The total monochromatic opacities
are tabulated and we also tabulate separately the absorption and scattering contributions, with the absorption contribution consisting of the first three terms of the right-hand-side of Eq.~(2) and
the scattering contribution as discussed in Section 2.5. We note that the Rosseland mean opacities available from our website
are computed assuming a refractive index $n_{\nu}$ set equal to 1.0 above the plasma frequency, and to zero below that frequency, 
in the integration that appears in Eq.~(1). 
This choice is made so that mixtures of opacities may be computed in a more straightforward manner. 
Mixtures of opacities are generated by mixing the pure-element OPLIB tables under the assumption of electron-temperature and electron-degeneracy equilibrium \citep{huebner}.

The user may choose to obtain opacities from the latest OPLIB tables (generated using ATOMIC) or from the previous set of OPLIB
tables (generated using LEDCOP).  The new ATOMIC-generated OPLIB tables are available on a more refined temperature grid, with 
24 more isotherms available compared to the grid on which the previous LEDCOP calculations were made. This should help
reduce interpolation errors that may arise when interpolating our opacity tables onto a different temperature grid. Such issues
were discussed in detail by \citet{guzik1}.
Also, the new opacity
tables generally extend to higher mass densities than were available from the OPLIB tables computed using LEDCOP. 
Finally, as a service to the community, opacity tables are available for a variety of pre-calculated mixtures
of elements.


\section{Results}

\subsection{Single-Element Opacity Comparisons}

In our previous publications we have performed several sets of comparisons of our new opacity calculations with measurement and
other theoretical opacity efforts that are available in the literature. For example, in \citet{colgan3} we compared monochromatic opacities of Al and Fe with 
various measurements performed in the 1990s \citep{foster,springer,perry,winhart}. Overall, good agreement was found with these
measurements. We have also compared Rosseland mean opacities for the elements H and He \citep{colgan1}, and C and O \citep{colgan2} 
with OP and OPAL data. Again, overall reasonable, although not perfect, agreement was found between the ATOMIC calculations
and other opacity efforts. In \citet{colgan3} we also examined the opacity of transition metal elements at the somewhat lower temperature
of 15.3~eV. We identified significant differences between our ATOMIC opacities and those generated using OP \citep{op2} calculations 
for Fe, Ni, and Cr. In particular, the systematic trend in the monochromatic opacity that is apparent in the ATOMIC calculations as one
moves from Cr to Fe to Ni is not observed in the OP calculations. The differences in the monochromatic opacities of these elements
lead to an increase of around a factor of two in the ATOMIC calculations of the Rosseland mean opacity compared to the OP calculations.

In figure~1 we examine the opacity of Mg at a temperature and density relevant to the base of the convection zone of the Sun, that is a
temperature of 192.91~eV and an electron density of $10^{23}$ cm$^{-3}$. This
monochromatic opacity was also examined by the OPAS team \citep{opas}. We compare the ATOMIC calculation to the OP calculations
(available online at {\it http://cdsweb.u-strasbg.fr/topbase}) and find reasonable agreement in the broad features of the monochromatic opacity. However, we do note that
the strong He$_{\alpha}$ line at a photon energy of 1352~eV is broader in the OP calculation than in the ATOMIC calculation.
In the ATOMIC calculation, the Stark broadening of this line is computed using the package of Lee \citep{starklee}, modified to utilize the
atomic data available from CATS. It is unclear as to how the broadening of the He-like lines is computed within the OP calculations.

The broader He$_{\alpha}$ line from the OP calculation was also noted by the work of \citet{opas}, and the OPAS opacity appears quite close to the ATOMIC opacity shown in figure~1. The inset of figure~1 shows the ionization balance of Mg for these conditions. Very close agreement is found between the OP, OPAS and
ATOMIC calculations. 

\begin{figure}
\epsscale{.80}
\plotone{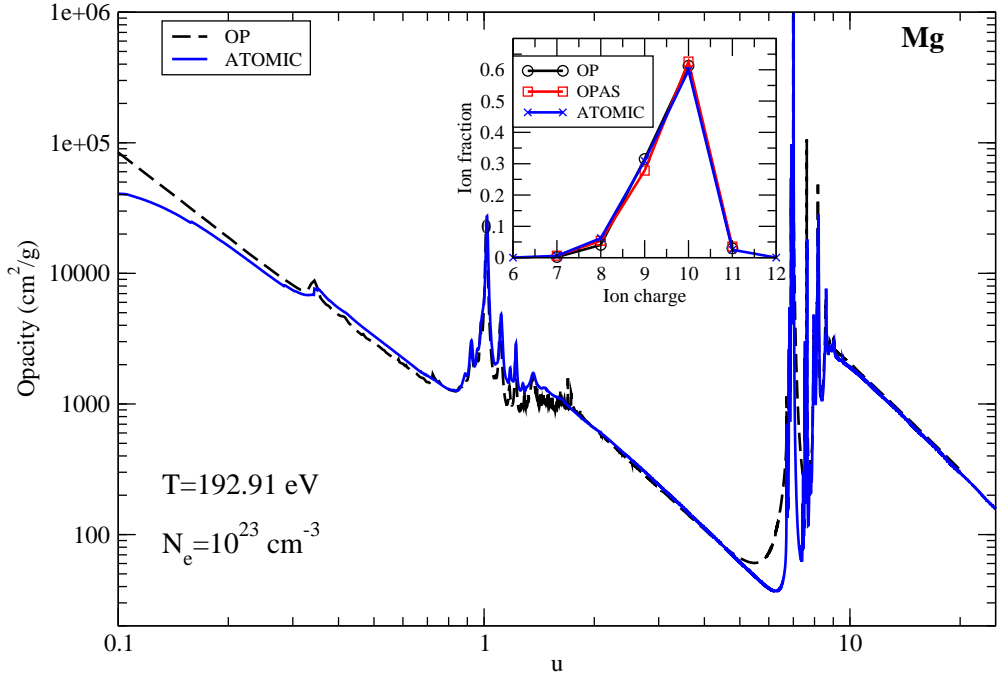}
\caption{ \label{fig1} Comparison of Mg opacity (as a function of $u=h\nu/kT$) at a temperature of 192.91 eV and an electron density of $10^{23}$ cm$^{-3}$. The current ATOMIC calculation (blue curve) is compared with an OP calculation (black dashed line) \citep{op2}. 
The inset shows the ionization balance of Mg at the same conditions where we compare
the current ATOMIC calculations (blue crosses) with OP (black circles) \citep{op2} and OPAS (red squares) calculations \citep{opas}. }
\end{figure}

In figure~2 we examine the opacity of Fe and compare our ATOMIC calculations to the measurements of \citet{bailey07} made using
the Sandia National Laboratory Z-pinch platform. We find very good agreement between our calculations and the 2007 measurements. 
There is excellent agreement between the measured and calculated line positions and valleys between the lines (which are important
in Rosseland mean opacity calculations), as well as in the underlying opacity, although some difference is observed in some of the
line heights. This may be due to an incomplete treatment
of the effects of instrument resolution in the calculations. We
note that similar calculations to the ones shown here were reported by \citet{bailey07} and were also made using the MUTA option in ATOMIC,
but with a different atomic data set. The calculations shown in figure~2 were made using the same atomic models that were used to construct
the Fe OPLIB table that is available through our opacity website.

\begin{figure}
\epsscale{.80}
\plotone{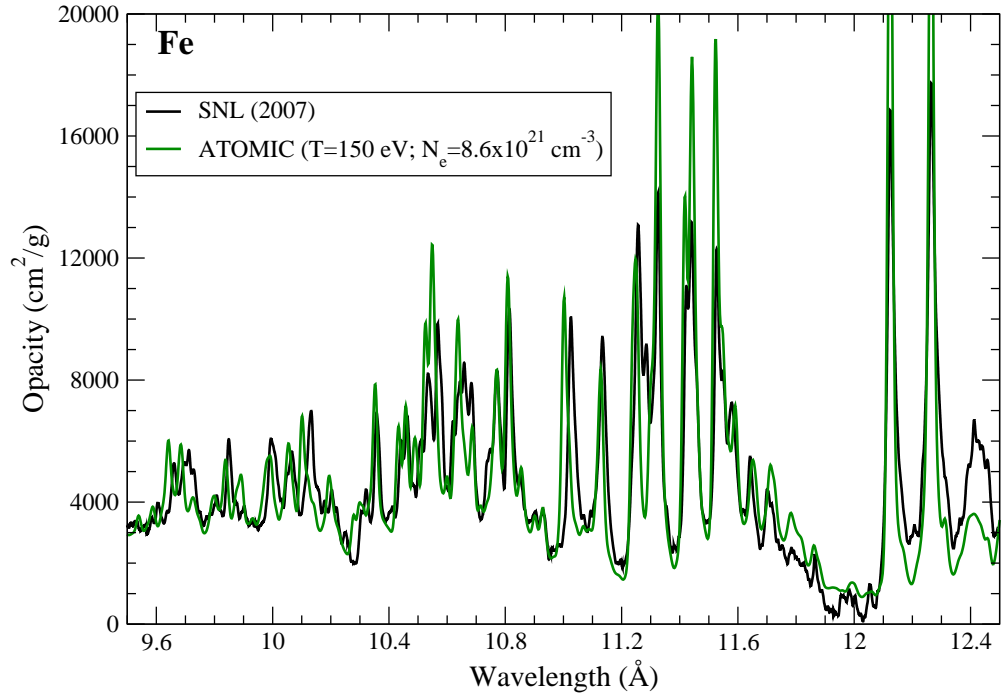}
\caption{ \label{fig2} Comparison of Fe opacity measured in 2007 \citep{bailey07} using the Sandia National Laboratory Z-pinch (black line) with an ATOMIC
calculation at a temperature ($T$) of 150 eV and an electron density ($N_e$) of $8.6\times10^{21}$ cm$^{-3}$. The experimental conditions
were inferred to be $T=156\pm6$ eV and $N_e=6.9\times10^{21}$ cm$^{-3}$ $\pm25\%$ \citep{bailey07}. }
\end{figure}


We now discuss the more recent experiments that measured the opacity of Fe using the refurbished Sandia Z-pinch platform \citep{bailey15}, which produces more energy per shot. 
This, coupled with a change in the design of the tampers of the Fe/Mg targets, enabled hotter temperatures and
higher densities to be explored. (The Fe target is combined with layers of Mg so that the Mg lines can be used as a diagnostic to obtain
estimates of the plasma electron density and temperature). The inferred conditions (from the Mg line diagnostics) implied a plasma
temperature of $182\pm7$~eV with an electron density of $3.1\pm0.78\times10^{22}$ cm$^{-3}$. However, ATOMIC calculations at those conditions
(as shown in figure~3 of \citet{bailey15}) are in poor agreement with the opacity inferred from the experiment. In particular, although ATOMIC is in good agreement with the positions
of the major line features, a persistent background discrepancy is found between measurement and calculation. We note especially the
disagreement at the lowest wavelengths, where the bound-free contribution (from the $L$-shell) to the opacity dominates. As pointed out in \citet{bailey15},
these measurements (and disagreements with theory), if confirmed, have important implications for solar opacities near the base of the radiative convection
zone.

The large disagreement between the ATOMIC calculations and the Sandia measurements prompted us to re-examine many aspects of our
calculations. In particular, the higher electron density inferred in the more recent measurements led to speculation \citep{bailey15} that much of the
population of the relevant Fe ion stages resided in excited states, and that this population may not be accurately portrayed in the calculations.  To test this hypothesis, we constructed atomic data sets comprised of larger numbers of multiply-excited configurations than used in our normal 
Fe opacity tables (which already
included a considerable number of configurations that represented multiply-excited-states). This was accomplished
by including promotions of four electrons from the $L$-shell of the relevant ions that were populated in this calculation (which ranged from N-like 
through Mg-like Fe).
The original ATOMIC calculations already included promotions of at least two electrons from the $L$-shell. This inclusion
resulted in calculations that included around one order of magnitude more configurations than in the calculations used originally. This
increased number of configurations was then used in the atomic structure calculations, and ATOMIC was again used
to compute the resulting opacity. This effort led to only small changes in the monochromatic opacity compared to the original ATOMIC calculations, and in particular
the bound-free opacity (which dominates the total opacity at lower wavelengths) was almost unchanged.  We also examined several other
aspects of our calculations. We tested the effects of inclusion of full configuration-interaction (CI) within our models by constructing a smaller
set of configurations for the relevant Fe ion stages for which a full CI calculation was feasible. The effects of full CI were found to be minor
and again had almost no change on the resulting bound-free cross sections. We also tested the sensitivity of our calculations to the
choice of screening microfield distribution in the occupation probability (as discussed in the previous section). Again, at these conditions, our calculations were not sensitive to the choice of screening within the microfield.

We conclude this discussion by noting that other modern opacity calculations also disagree with the Sandia measurements, as discussed
in \citet{bailey15}. We also note a recent paper \citep{iglesias15a} that implies that the Sandia measurements, if correct, 
are in apparent violation of oscillator strength sum rules.
%
Another important related issue is related to the bound-free opacity of the Fe plasma at these conditions. If this
is indeed underestimated in the calculations, this has serious implications for long-established methods in atomic physics for computing photoionization cross
sections from highly-charged ions. Our approach to photoionization uses a distorted-wave approach, which is usually thought to be
of acceptable accuracy for photoionization calculations from moderately- and highly-charged systems.
On the other hand, the measurements reported in 
\citet{bailey15} have been subject to considerable scrutiny \citep{naga14} in the search for experimental issues that might affect the measured
opacity. No significant systematic errors that could artificially increase the measured opacity were reported by \citet{naga14}.
We look forward to independent verification of these measurements
and to a resolution of this discrepancy.


\begin{figure}
\epsscale{.80}
\plotone{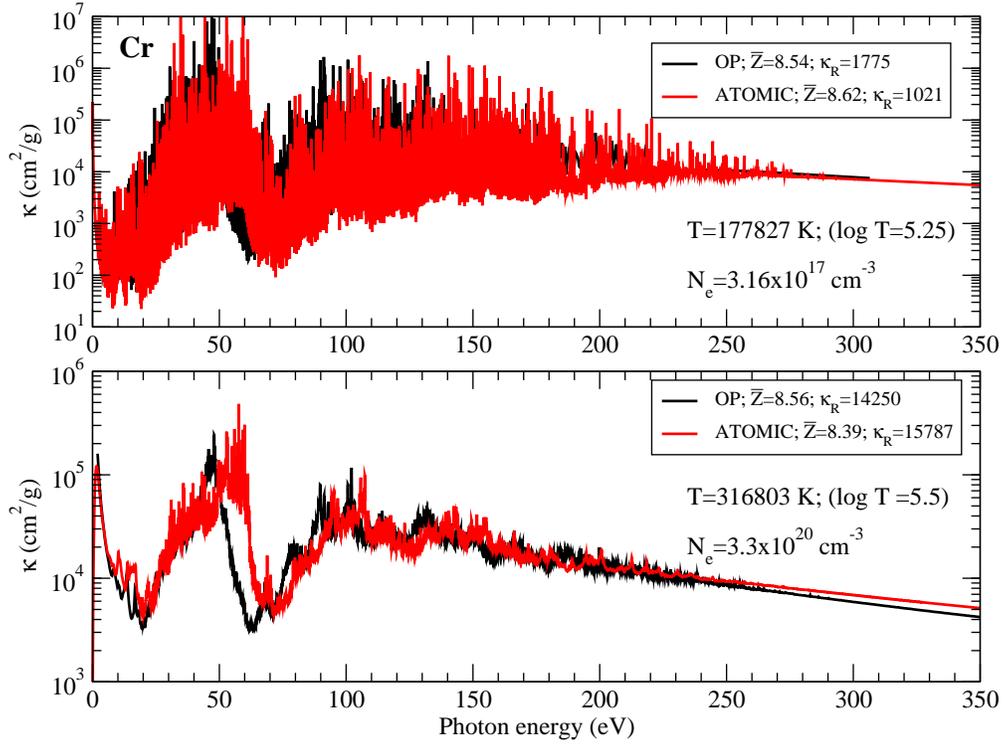}
\caption{ \label{fig3} Comparison of the opacity of Cr from the OP database \citep{op2} (black lines) and the current ATOMIC calculations (red lines) at two different sets of temperatures ($T$) and electron densities ($N_e$) as indicated. The average ionization (${\overline Z}$), and Rosseland
mean opacity ($\kappa_{\rm R}$), in cm$^2$/g, are indicated in the figure captions. }
\end{figure}	

We now turn to a discussion of opacities at conditions relevant to B stars. The pulsation properties of such stars were recently
explored \citep{walczak15} using our new Los Alamos opacities. Previous work \citep{gilles11,stc13} has cast some doubt on the use of the OP database for such systems. In particular, the Ni OP opacities, which made use of scaled atomic data \citep{op2} were found to be in significant
disagreement with several other sets of opacity calculations \citep{stc13}. Studies of Fe opacities at these conditions, including
previous ATOMIC calculations \citep{colgan1}, also indicated that several important inner-shell transitions may have been omitted from the
OP calculations, a conclusion also reached by \citet{iglesias15b}.

Since the differences between the OP database and more recent calculations for Fe and Ni have been documented, we here examine
the opacity of Cr at conditions relevant for B stars. Figure~3 shows a comparison of the OP calculations and ATOMIC calculations of Cr
at two sets of conditions. The upper panel shows the opacity at a temperature corresponding to log($T$)=5.25 (temperature $T$ in K) and an electron density of $3.16\times 10^{17}$ cm$^{-3}$; this corresponds to conditions near the opacity bump ($Z$-bump) that is evident when the Rosseland mean opacity is plotted as a function of temperature for constant log~$R$ values (where $R = \frac{\rho}{T_6^3}$, with $\rho$ the mass density in g/cm$^3$ and $T_6=10^{-6}T$).
We find that both the OP and ATOMIC calculations produce monochromatic opacities that exhibit a dense forest of lines due to the very large
number of bound-bound transitions that contribute to the total opacity at these conditions. Although it is difficult to make a meaningful 
comparison of the two calculations for such a dense spectrum, it does appear that the ATOMIC opacity is somewhat shifted to higher
photon energies compared to the OP opacity. Furthermore, we note that the Rosseland mean opacity from the OP calculation is
considerably higher than that from the ATOMIC calculation. In an effort to explore this further, in the lower panel of figure~3 we examine the Cr opacity at a higher temperature and electron density, following a strategy proposed by \citet{stc13}. 
This set of conditions was chosen to produce a similar ionization balance to the 
conditions in the upper panel. 
This is reflected by the reasonably similar average ionization (${\overline Z}$) found for these conditions
as indicated in the figure. The monochromatic opacity in the lower panel has features that are much broader than the narrow lines
evident in the upper panel due to the larger electron density in this case, which causes the bound-bound features to broaden and merge
with each other.
Although for the conditions in the lower panel we find that the Rosseland mean opacities from OP and ATOMIC 
are closer than in the upper panel, we note that the ATOMIC calculations are again shifted to higher photon energies compared to the
OP calculations. Since we remember that the OP calculation for Cr (as well as for Ni and Mn) used scaled atomic data \citep{op2}, this
shift may be due to this approximation. This is partially confirmed by comparison to LEDCOP calculations (not shown), which are
in very good agreement with the ATOMIC calculations. 
%

\subsection{Solar Model Results}

We calculated standard solar evolution models using the ATOMIC and OPAL opacities for the \citet{asplund09} (AGSS09)
photospheric abundance mixture (see \citet{Guzik2015a,Guzik2015b}).  The ATOMIC opacity tables were generated by mixing the pure-element OPLIB tables under the assumption of electron-temperature and electron-degeneracy equilibrium.

\begin{figure}
\begin{center}
\epsscale{.80}
\plotone{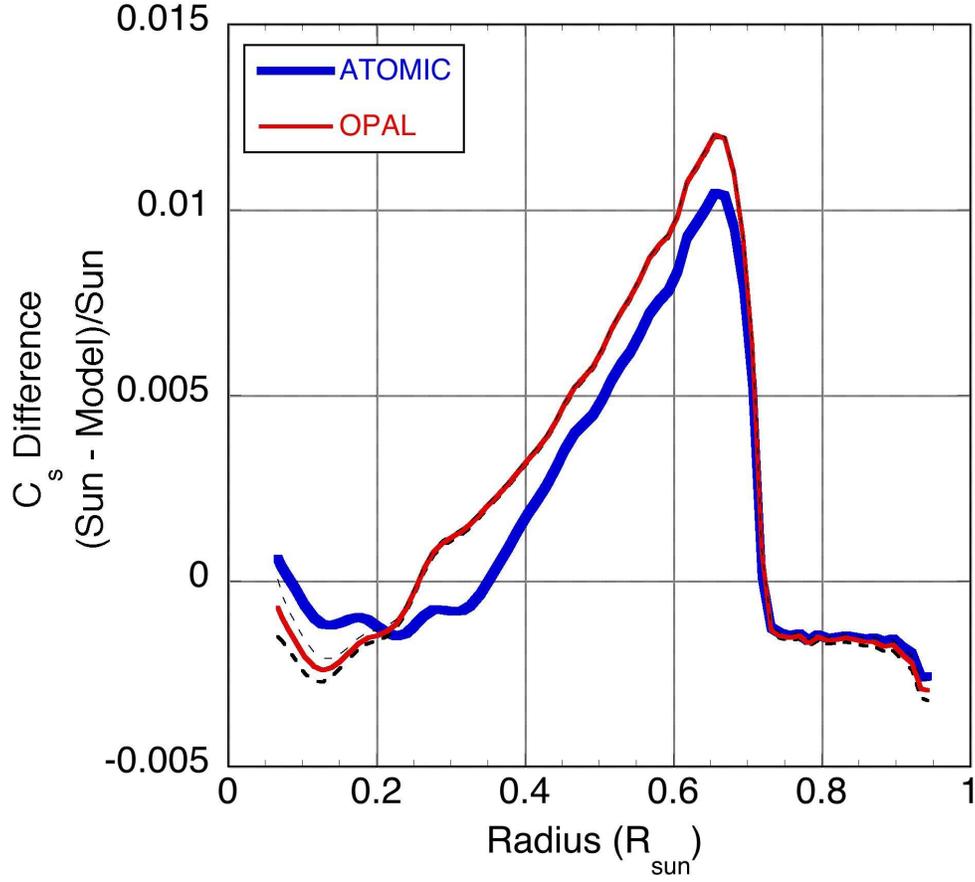}
 \caption{Helioseismically inferred \citep{Basu2000} minus calculated sound speed  differences vs. radius for solar models using the ATOMIC and OPAL opacities with the AGSS09 abundance mixture.  The black dashed curves on either side of the OPAL profile show the magnitude of the uncertainty in the inferred sound-speed profile. The convection zone base radius is at $\sim$ 0.725 R$_{\odot}$.}
\label{soundspeed}
\end{center}
\end{figure}

\begin{figure}
\begin{center}
\epsscale{.80}
\plotone{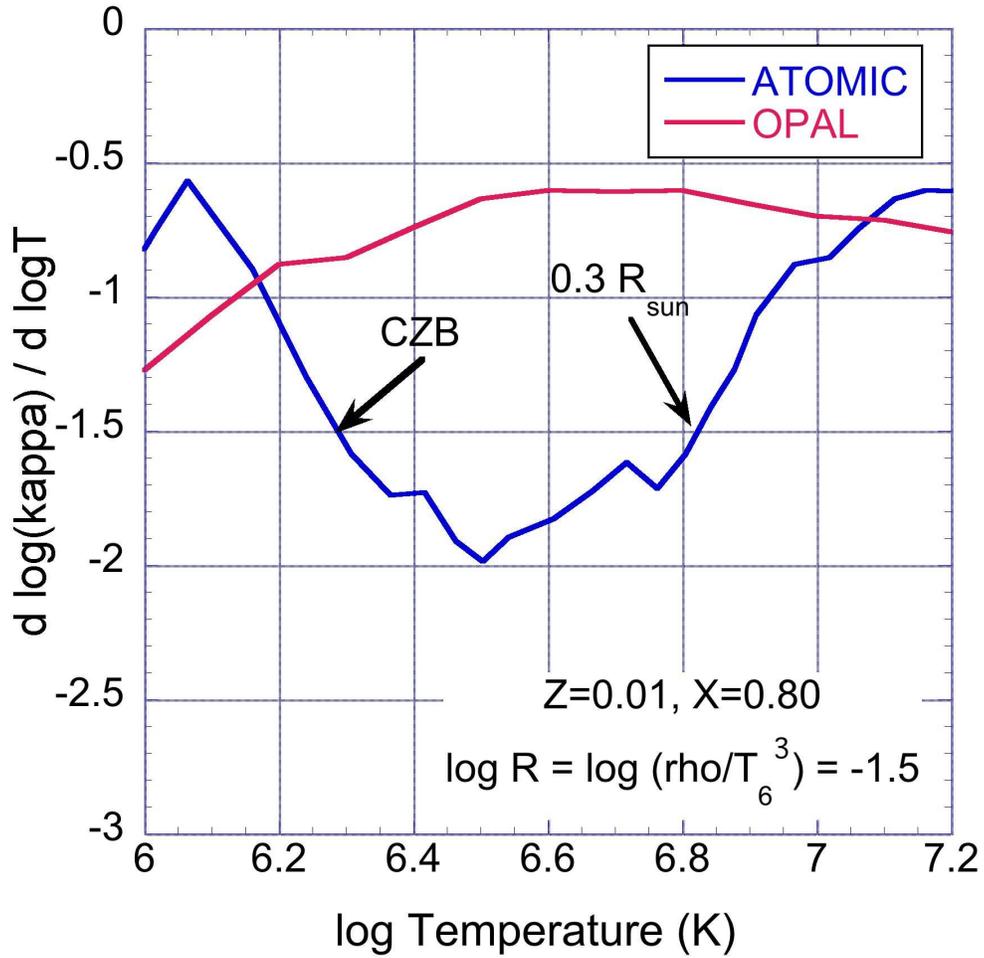}
 \caption{Logarithmic opacity derivative with respect to temperature for ATOMIC and OPAL tables with Z=0.01, X=0.80, and log $R$ = -1.5.  The larger gradients for the ATOMIC table opacities at the convection-zone base (log $T$ = 6.3) and at $\sim$0.2-0.3 R$_\odot$ (log T $\sim$6.8)
 in the solar interior are responsible for the differences in sound-speed gradients between models seen in Fig. \ref{soundspeed}.}
\label{gradient}
\end{center}
\end{figure}
 
\begin{table}
  \begin{center}
  \caption{Calibration parameters and properties of solar models evolved using OPAL or ATOMIC opacities for the AGSS09 abundance mixture.}
  \label{tab1}
  \begin{tabular}{lcc}\hline
 & {\bf OPAL} & {\bf ATOMIC} \\
 \hline
{\bf Y$_{initial}$}  & 0.2641 & 0.2570 \\
{\bf Z$_{initial}$}  & 0.0150 & 0.0151 \\
{\bf $\alpha$ }  & 2.0118 & 2.0637 \\
\\
{\bf Y$_{conv. zone}$$^a$}  & 0.2345 & 0.2283 \\
{\bf Z$_{conv. zone}$}  & 0.0135 & 0.0136 \\
{\bf R$_{conv. zone~base}$$^b$ (R$_{\odot}$)} & 0.7264 & 0.7251 \\
 \hline
  \end{tabular}
 \end{center}
 $^a$Helioseismically inferred convection-zone Y is 0.248 $\pm$ 0.003 \citep{BasuAntia2004}\\
 $^b$Helioseismically inferred convection-zone radius is 0.713 $\pm$ 0.001 R$_{\odot}$ \citep{BasuAntia2004}
\end{table}
 
The solar models were calculated using an updated version of the Iben evolution code (see \citet{guzik10} for details).  The models include diffusive settling of helium and heavier elements relative to hydrogen.  We adopt the usual symbols for hydrogen mass fraction (X), helium mass fraction (Y), and mass fraction of all elements heavier than hydrogen and helium (Z), such that X+Y+Z=1. The initial helium mass fraction and mixing length to pressure-scale-height ratio ($\alpha$) are adjusted to calibrate the model to the observed solar luminosity 
($3.846\times10^{33}$ erg s$^{-1}$) and radius ($6.9599\times10^{10}$ cm) at the present solar age (4.54 $\pm$ 0.04 billion years).   The initial Z is also adjusted so that, after diffusive settling, at the present solar age, the photospheric Z/X = 0.0181, in agreement with the value derived by AGSS09.  Table \ref{tab1} summarizes the calibration parameters and other properties of the two solar models.  The model evolved with the ATOMIC opacities has a slightly deeper convection zone than the model evolved with the OPAL opacities, but  both models still show a too-shallow convection-zone depth and too-low convection zone helium mass fraction compared to the helioseismically inferred values from \citet{BasuAntia2004}.

Figure \ref{soundspeed} shows helioseismically inferred \citep{Basu2000} minus calculated sound speed vs. radius for the two models.  The calculated sound speed profile is in better agreement with helioseismic inference using the ATOMIC opacities, although this change alone does not resolve the discrepancy.  We also found a similar improvement in agreement of sound speed profile with helioseismic inferences for solar models evolved using the AGSS09 abundance mixture calculated using the MESA code (\citep{MESA}, see also \citet{Guzik2015a,Guzik2015b}).
 
To investigate the reason for the change in sound speed profile using the ATOMIC vs. the OPAL opacities, we compared the absolute values and the logarithmic temperature derivatives of the OPAL and ATOMIC opacities.  We find that the OPAL opacities are actually slightly higher than the ATOMIC opacities for the entire solar radiative interior (log $T$ = 6.3 to 7.2)  but that the ATOMIC opacities become higher than OPAL in the solar convection zone, where opacity is not important to the solar structure because convection is transporting nearly all of the emergent luminosity.  Figure \ref{gradient} shows the logarithmic opacity derivatives with respect to temperature for the ATOMIC and OPAL table opacities for log $R$ = -1.5, Z=0.01, and X=0.8.  The derivatives of the ATOMIC opacities are steeper than those of the OPAL opacities at the location of the solar convection zone base around 0.725 R$_{\odot}$ (log $T$ $\sim$6.3), and near log $T$ = 6.8, corresponding to 
$\sim$0.3 R$_{\odot}$.  These are the locations that the sound-speed profile for the solar model using the ATOMIC opacities changes slope compared to the model for the OPAL opacities (Fig. \ref{soundspeed}).

\section{Conclusions}

In this paper we have discussed in detail the calculations of the new OPLIB opacity tables that have been generated
using the Los Alamos ATOMIC code. Our calculations represent a systematic improvement in both the underlying physics approximations and in the amount of detail included in the calculations compared to the previous generation of
Los Alamos opacity tables that were released around 15 years ago. 
A selection of our new monochromatic opacities have been discussed here and also in recent publications \citep{colgan1,colgan2,colgan3}, where we presented detailed comparisons of ATOMIC and LEDCOP calculations. 
We have also demonstrated that use of our new opacity tables in solar modeling leads to 
improved agreement with helioseismology, although use of the new tables does not fully resolve the long-standing discrepancies. We also note
that our new tables lead to improved agreement with observation of pulsations of B-stars \citep{walczak15}.

We hope that our new opacity tables, which are now publicly available\footnote{http://aphysics2.lanl.gov/opacity/lanl}, will prove useful to the
astrophysical modeling community. Further exploration of how these new opacities impact the understanding of a multitude of stellar objects
is highly desirable and will be investigated in the near future.

\acknowledgments

The Los Alamos National Laboratory is operated by Los Alamos
National Security, LLC for the National Nuclear Security
Administration of the U.S. Department of Energy under Contract No.
DE-AC52-06NA25396.
We thank P. Walczak, who helped prepare the ATOMIC and OPAL tables for the solar model results presented here.
We obtained LLNL opacities from the Lawrence Livermore National Laboratory OPAL Opacity Web site:  http://opalopacity.llnl.gov/opal.html.

\section{Appendix: Rayleigh scattering details}

We approximate the low energy Rayleigh scattering contribution for all neutral atoms based on the static dipole polarizability of the neutral atom \citep{C76}. When more detailed treatments are available they are used in place of this approximation (currently for H, He, Li, C, N, O, Ne, and Ar).  The more detailed treatments are based on the dynamic (or frequency dependent) dipole polarizability \citep{TV69}. The following formulae have been implemented in ATOMIC for the scattering on the red wing of the first resonance line of the ground state of the neutral atom. Below
we give expressions for the scattering transport cross section $\sigma$,
where also $\hbar\omega$ is the photon energy (in eV),
$E_H= 13.6057$~eV is the Rydberg unit of energy and $\sigma_{Th}=  8\pi r_0^2/3 = 6.65246\times10^{-25}$ cm$^2$ is the total Thomson scattering cross section.

{\it Hydrogen}: \\
From \citet{Lee05} we have
\begin{equation}
\sigma/\sigma_{Th} = 20.24 \left(\frac{\hbar\omega}{2 E_H} \right)^4 + 239.2 \left(\frac{\hbar\omega}{2 E_H} \right)^6 + 2256 \left(\frac{\hbar\omega}{2 E_H} \right)^8 .
\end{equation}

{\it Helium}: \\
From \citet{Dal62, Dal60}, but rescaled so that the first term agrees with more accurate static dipole polarizability factors from \citet{nb}, we have
\begin{equation}
\sigma/\sigma_{Th} = 1.913 \left(\frac{\hbar\omega}{2 E_H} \right)^4 + 4.52 \left(\frac{\hbar\omega}{2 E_H} \right)^6 + 7.90 \left(\frac{\hbar\omega}{2 E_H} \right)^8 .
\end{equation}

{\it Lithium}:  \\
From \citet{Zeiss77}, using the expansion of the dynamic polarizability given by
\begin{equation}
\alpha(\omega) = \frac{e^2\hbar^2}{m}\sum_{j = 1}^\infty (\hbar\omega)^{2j-2} S(-2j) \ ,
\label{CM1}
\end{equation}
where the individual Cauchy moments are defined  by
\begin{equation}
S(-2j) = \sum_N \frac{f_{N,I}}{(E_{N,I})^{2j}} \  ,
\label{CM2}
\end{equation}
we obtain
\begin{equation}
\sigma/\sigma_{Th} =  \left(\frac{\hbar\omega}{2 E_H} \right)^4 \left(163.6 + 35038 \left(\frac{\hbar\omega}{2 E_H} \right)^2 + 7.590\times10^6\left(\frac{\hbar\omega}{2 E_H} \right)^4 \right)^2 .
\end{equation}

{\it Carbon}:  \\
From \citet{TV69}, but rescaled so that the first term agrees with more accurate static dipole polarizability factors from \citet{nb}, we have
\begin{equation}
\sigma/\sigma_{Th} = 126.8 \left(\frac{\hbar\omega}{2 E_H} \right)^4 \left(  1 +  12.87 \left(\frac{\hbar\omega}{2 E_H} \right)^2 + 152.3 \left(\frac{\hbar\omega}{2 E_H} \right)^4 \right) .
\end{equation}

{\it Nitrogen}:  \\
From \citet{TV69}, but rescaled so that the first term agrees with more accurate static dipole polarizability factors from \citet{nb}, we have
\begin{equation}
\sigma/\sigma_{Th} = 54.91 \left(\frac{\hbar\omega}{2 E_H} \right)^4 \left(  1 +  9.611 \left(\frac{\hbar\omega}{2 E_H} \right)^2 + 78.03 \left(\frac{\hbar\omega}{2 E_H} \right)^4 \right) .
\end{equation}

{\it Oxygen}:  \\
From \citet{TV69}, but rescaled so that the first term agrees with more accurate static dipole polarizability factors from \citet{nb}, we have
\begin{equation}
\sigma/\sigma_{Th} =  36.63 \left(\frac{\hbar\omega}{2 E_H} \right)^4 \left(  1 +  4.803 \left(\frac{\hbar\omega}{2 E_H} \right)^2 + 23.44 \left(\frac{\hbar\omega}{2 E_H} \right)^4 \right) .
\end{equation}

{\it Neon}:  \\
From \citet{Dal62, Dal60}, but rescaled so that the first term agrees with more accurate static dipole polarizability factors from \citet{nb}, we have
\begin{equation}
\sigma/\sigma_{Th} = 7.129  \left(\frac{\hbar\omega}{2 E_H} \right)^4 \left(  1 +  2.16 \left(\frac{\hbar\omega}{2 E_H} \right)^2 + 4.92 \left(\frac{\hbar\omega}{2 E_H} \right)^4 \right) .
\end{equation}

{\it Argon}:  \\
From \citet{Dal62, Dal60}, but rescaled so that the first term agrees with more accurate static dipole polarizability factors from \citet{nb}, we have
\begin{equation}
\sigma/\sigma_{Th} =   122.55\left(\frac{\hbar\omega}{2 E_H} \right)^4 \left(  1 +   2.48\left(\frac{\hbar\omega}{2 E_H} \right)^2 +  9.72\left(\frac{\hbar\omega}{2 E_H} \right)^4 \right) .
\end{equation}

{\it Other Elements}:  \\
When more complete expressions are not available, we use the static dipole polarizability approximation 
using 
$\alpha(\omega=0)$ to give the total cross section for scattering opacity as
\begin{equation}
\sigma/\sigma_{Th} = \alpha(0)^2 \left(\frac{\hbar\omega}{2 E_H} \right)^4 .
\end{equation}
In the final expression given above, the dipole polarizability $\alpha(0)$ is in units of Bohr radii cubed.

Finally, the scattering opacity contribution, $\kappa_{\rm SCAT}$, is given by
\begin{equation}
\kappa_{\rm SCAT} = \frac{N \sigma}{\rho} \ ,
\end{equation}
with $\rho$ the mass density in g/cm$^3$, and $N$ the population of the ground state (in cm$^{-3}$) of the neutral atom under consideration.


%



\begin{thebibliography}{}

%
%
\bibitem[Abdallah et al.(1988)]{cats} Abdallah, Jr., J., Clark, R. E. H., \& Cowan, R. D. 1988,
Los Alamos Manual LA-11436-M, Vol.~I

\bibitem[Abdallah et al.(2007)]{abd07} Abdallah, Jr., J., Kilcrease, D. P., Magee, N. H., Mazevet, S., Hakel, P., \& Sherrill, M. E. 2007,
High Energy Density Phys., 3, 309

\bibitem[Ali et al.(1965)]{ali1} Ali, A. W., \& Griem, H. R. 1965, Phys. Rev., 140, A1044

\bibitem[Ali et al.(1966)]{ali2} Ali, A. W., \& Griem, H. R. 1966, Phys. Rev., 144, 366

\bibitem[Armstrong et al.(1966)]{arm1} Armstrong, B. H., Johnson, R. R., DeWitt, H. E., \& Brush, S. G. 1966, Opacity
of High Temperature Air, ed. C. A. Rouse (Progress in High
Temperature Physics and Chemistry, Vol 1., Pergamon Press, Oxford,
New York)

\bibitem[Armstrong et al.(2014)]{armstrong} Armstrong, G. S. J., Colgan, J., Kilcrease, D. P., \& Magee Jr., N. H. 2014, High Energy Density Phys., 10, 61

\bibitem[Asplund et al.(2005)]{asplund05} Asplund, M., Grevesse, N., \& Sauval, A. J. 2005, in ASP Conf. Ser., 336, 25

\bibitem[Asplund et al.(2009)]{asplund09} Asplund, M., Grevesse, N., Sauval, A. J., \& Scott, P. 2009, ARA\&A, 47, 481

\bibitem[Badnell et al.(2005)]{op2} Badnell, N. R., Bautista, M. A., Butler, K., Delahaye, F., Mendoza, C., Palmeri, P., Zeippen, C. J.,
\& Seaton, M. J. 2005, MNRAS, 360, 458

\bibitem[Bahcall et al.(2005)]{bahcall05} Bahcall, J. N., Serenelli, A. M., \& Basu, S. 2005, ApJ, 621, L85


\bibitem[Bailey et al.(2007)]{bailey07} Bailey, J. E., Rochau, G. A., Iglesias, C. A., Abdallah Jr., J., 
MacFarlane, J. J., Golovkin,~I., Wang,~P., Mancini,~R.~C., Lake, P. W., Moore, T. C., Bump, M., Garcia, O., \& Mazevet, S. 
2007, Phys. Rev. Letts., 99, 265002

\bibitem[Bailey et al.(2015)]{bailey15} Bailey, J. E., Nagayama, T., Loisel, G. P., Rochau, G. A., Blancard, C., Colgan, J., Cosse, Ph., Faussurier, G., Fontes, C. J., Gilleron, F.,  Golovkin, I., Hansen, S. B., Iglesias, C. A., Kilcrease, D. P., MacFarlane, J. J., Mancini, R. C., Orban, C., Pain, J.-C., Pradhan, A. K., Sherrill, M., \& Wilson, B. G. 2015, Nature, 517, 56


\bibitem[Basu et~al.(2000)]{Basu2000}{Basu, S. et al.} 2000, ApJ, 529, 1084
 
\bibitem[Basu \& Antia(2004)]{BasuAntia2004}{Basu, S. \& Antia, H. M.} 2004, ApJ, 606, L85

\bibitem[Bell et al.(1988)]{bell} Bell, K. L., Hibbert, A., \& Berrington, K. A. 1988, J. Phys. B, 21, 2319

\bibitem[Blancard et al.(2012)]{opas} Blancard, C., Coss\'e, P., \& Faussurier, G. 2012, ApJ, 745, 10

\bibitem[Carson(1976)]{C76} Carson, T. R. 1976, Ann. Rev. Ast. Ap., 14, 95


\bibitem[Chabrier \& Potekhin(1998)]{cp98} Chabrier, G. \& Potekhin, A. Y. 1998, Phys. Rev. E, 58, 4941

\bibitem[Clark et al.(1991)]{gipper} Clark, R. E. H., Abdallah Jr, J., \& Mann, J. B. 1991,  ApJ, 381, 597;
Abdallah~Jr, J., Zhang, H.~L., Fontes, C.~J.,  Kilcrease, D.~P., \& Archer, B. J. 2001,
Journal of Quant. Spectrosc. Rad. Transfer, 71, 107

\bibitem[Colgan et al.(2013a)]{colgan1} Colgan, J., Kilcrease, D. P., Magee, Jr., N. H., Armstrong, G. S. J., Abdallah, Jr., J., Sherrill, M. E.,
Fontes, C. J., Zhang, H. L., \& Hakel, P. 2013a,
Proceedings of the International Conference on Atomic and Molecular Data,  pp~17

\bibitem[Colgan et al.(2013b)]{colgan2} Colgan, J., Kilcrease, D. P., Magee, Jr., N. H., Armstrong, G. S. J., Abdallah, Jr., J., Sherrill,~M.~E.,
Fontes,~C.~J., Zhang, H. L., \& Hakel, P. 2013b, High Energy Density Phys., 9, 369

\bibitem[Colgan et al.(2015)]{colgan3} Colgan, J., Kilcrease, D. P., Magee, Jr., N. H., Abdallah, Jr., J., Sherrill,~M.~E.,
Fontes,~C.~J., Zhang,~H.~L., \& Hakel, P. 2015, High Energy Density Phys., 14, 33

\bibitem[Cowan(1981)]{cowan} Cowan, R. D. 1981, The Theory of Atomic Structure and Spectra (University of
California Press, Berkeley)

\bibitem[Cugier(2012)]{cugier} Cugier, H. 2012, A \& A, 42, 547


\bibitem[Dalgarno(1962)]{Dal62} Dalgarno, A. 1962, Spectral Reflectivity of the Earth's Atmosphere III: The Scattering of Light by Atomic Systems, Geophysical Corporation of America Report No. 62-20-A

\bibitem[Dalgarno \& Kingston(1960)]{Dal60} Dalgarno, A., \& Kingston, A. E. 1960, Proc. Royal Soc. London A, 259, 424



\bibitem[D\"appen(1987)]{dappen87} D\"appen, W. 1987, ApJ, 319, 195

\bibitem[Daszy\'nska-Daszkiewicz \& Walczak(2009)]{dd09} Daszy\'nska-Daszkiewicz, J., \& Walczak, P. 2009, MNRAS, 398, 1961
\bibitem[Daszy\'nska-Daszkiewicz \& Walczak(2010)]{dd10} Daszy\'nska-Daszkiewicz, J., \& Walczak, P. 2010, MNRAS, 403, 496
\bibitem[Daszy\'nska-Daszkiewicz \& et al.(2013)]{dd13} Daszy\'nska-Daszkiewicz, J. Szewczuk, W., \& Walczak, P. 2013, MNRAS, 431, 3396


\bibitem[Foster et al.(1988)]{foster} Davidson, S. J., Foster, J. M., Smith, C. C., Warburton, K. A., \& Rose, S. J. 1988, Appl. Phys. Letters, 52, 847

\bibitem[Fontes et al.(2015)]{fontes15} Fontes, C. J., Zhang, H. L., Abdallah Jr, J., Clark, R. E. H., Kilcrease, D. P., Colgan, J., Cunningham, R. T., Hakel, P., Magee N. H., \& Sherrill, M. E. 2015,
J. Phys. B, 48, 144014

\bibitem[Fontes et al.(2016)]{fontes16} Fontes, C. J., Colgan, J., \& Abdallah, Jr., J. 2016, Self-Consistent Large-Scale Collisional-Radiative Modeling, in Modern Methods in Collisional-Radiative Modeling of Plasmas, ed. by Yu. Ralchenko (Springer, New York), in press

\bibitem[Geltman(1962)]{gelt62} Geltman, S. 1962, ApJ, 136, 935

\bibitem[Geltman(1965)]{gelt65} Geltman, S. 1965, ApJ, 141, 376

\bibitem[Gilles et al.(2011)]{gilles11} Gilles, D.,Turck-Chi\`eze, S., Loisel, G., Piau, L., Ducret, J. E., Poirier, M., Blenski, T., Thais, F., Blancard, C., Coss\'e, P., Faussurier, G., Gilleron, F., Pain, J. C., Porcherot, Q., Guzik, J. A., Kilcrease, D. P., Magee, N. H., Harris, J., Busquet, M., Delahaye, F., Zeippen, C. J., \& Bastiani-Ceccotti, S. 2011,
High Energy Density Phys., 7, 312

\bibitem[Green(1958)]{green58}  Green, J. M. 1958, RAND Corporation Report RM-2223-AEC

\bibitem[Green(1960)]{green60}  Green, J. M. 1960, RAND Corporation Report RM-2580-AEC

\bibitem[Grevesse \& Noels(1993)]{gn93} Grevesse, N., \& Noels, A. 1993, in Origin and Evolution of the Elements, 15

\bibitem[Grevesse \& Sauval(1998)]{gs98} Grevesse, N. \& Sauval, A. J. 1998,                          
      Space Science Reviews 85, 161

\bibitem[Griem(1974)]{griem_book} Griem, H. R. 1974, Spectral Line Broadening by Plasmas (Academic Press, New York
and London)

\bibitem[Guzik \& Mussack(2010)]{guzik10} Guzik, J. A., \& Mussack, K. 2010, ApJ, 713, 1108


\bibitem[Guzik et~al.(2015a)]{Guzik2015a}{Guzik, J. A., Fontes, C. J., Walczak, P., Wood, S. R., Mussack, K., \& Farag, E.} 2015a, Proc. International Astronomical Union Focus Meeting 17:  Advances in Stellar Physics from Asteroseismology, Honolulu, HI, August 10-14, 2015, eds. C.S. Jeffery, J. A. Guzik, \& K. Pollard, Cambridge U. Press, in press.

\bibitem[Guzik et~al.(2015b)]{Guzik2015b}{Guzik, J. A., et al.} 2015b, in preparation.  

\bibitem[Hakel \& Kilcrease(2004)]{chemeos1} Hakel, P., \& Kilcrease, D. P. 2004, 14th Topical Conference on Atomic Processes in Plasmas,
Eds: J. S. Cohen, S. Mazevet, and D. P. Kilcrease, (New York: AIP), pp~190

\bibitem[Hakel et al.(2006)]{hakel} Hakel, P.,  Sherrill, M. E., Mazevet, S., Abdallah Jr., J., Colgan, J.,
Kilcrease, D. P., Magee, N. H., Fontes, C. J., \& Zhang, H. L. 2006,
J. Quant. Spectrosc. Rad. Transfer, 99, 265

\bibitem[Heitler(2010)]{heitler} Heitler, W. 2010, The Quantum Theory of Radiation (3rd Edition, Dover Publications)

\bibitem[Hindmarsh et al.(1967)]{vdw} Hindmarsh, W. R., Petford, A. D., \& Smith, G. 1967, Proc. Roy. Soc. A, 297, 296

\bibitem[Hubeny et al.(1994)]{hub94} Hubeny, I., et al. 1994, A \& A, 282, 151

\bibitem[Huebner \& Barfield(2014)]{huebner} Huebner, W. F., \& Barfield, W. D. 2014, Opacity (Springer, New York)

\bibitem[Hummer \& Mihalas(1988)]{hm88} Hummer, D., \& Mihalas, D. 1988, ApJ, 331, 794



\bibitem[Iglesias(2010)]{iglesias10} Iglesias, C. 2010, High Energy Density Phys., 6, 311

\bibitem[Iglesias(2015a)]{iglesias15a} Iglesias, C. 2015a, High Energy Density Phys., 15, 4

\bibitem[Iglesias(2015b)]{iglesias15b} Iglesias, C. 2015b, MNRAS, 450, 2

\bibitem[Iglesias and Rogers(1996)]{opal2} Iglesias, C. A., \& Rogers, F. J. 1996, ApJ, 464, 943

\bibitem[Kilcrease et al.(2015)]{chemeos2}  Kilcrease, D. P., Colgan, J., Hakel, P., Fontes, C. J., \& Sherrill, M. E. 2015,
High Energy Density Phys., 16, 36

\bibitem[Kilcrease \& Magee(2001)]{pb_paper} Kilcrease, D. P., \& Magee, N. H. 2001, J. Quant. Spectrosc. Radiat. Transfer, 71, 445

\bibitem[Lee(1988)]{starklee} Lee, R. W. 1988, J. Quant. Spectrosc. Radiat. Transfer, 40, 561

\bibitem[Lee(2005)]{Lee05} Lee, H. W. 2005, MNRAS, 358, 1472

{\color{red}
\bibitem[Le Pennec et al.(2015)]{lepennec2015} Le Pennec, M., Turck-Chi\`eze, S., Salmon, S., Blancard, C., Coss\'e, P., Faussurier, G., \& Mondet, G. 2015, ApJ, 813, L42}

\bibitem[Magee et al.(1995)]{ledcop} Magee, N. H., Abdallah, Jr., J., Clark, R. E. H., Cohen, J. S., Collins, L.~A., Csanak, G., Fontes,~C.~J., Gauger, A. Keady, J.~J., Kilcrease, D.~P., \& Merts, A.~L. 1995, Astronomical Society of the Pacific Conference Series, 78, 51

\bibitem[Magee et al.(2004)]{magee} Magee, N. H., Abdallah, J., Colgan, J.,  Hakel, P.,  Kilcrease, D.~P.,  Mazevet, S., Sherrill, M.,
Fontes,~C.~J.,  \& Zhang, H.~L. 2004, {14th Topical Conference on Atomic Processes in Plasmas},
Eds: J. S. Cohen, S. Mazevet, and D. P. Kilcrease, (New York: AIP), pp~168

\bibitem[Mazevet \& Abdallah(2006)]{ma} Mazevet S., \& Abdallah, Jr., J. 2006, J. Phys. B, 39, 3419

{\color{red}
\bibitem[Mondet et al.(2015)]{mondet} Mondet, G., Blancard, C., Coss\'e, P., \& Faussurier, G. 2015, ApJ SS, 220, 2 }

\bibitem[Nagayama et al.(2014)]{naga14} Nagayama, T., Bailey, J. E., Loisel, G., Hansen, S. B., Rochau, G. A., Mancini,~R.~C., MacFarlane,~J.~J., \& Golovkin, I. 2014, Phys. Plasmas, 21, 056502

\bibitem[Nakagawa et al.(1987)]{itoh} Nakagawa, M., Kohyama, Y., \& Itoh, N. 1987, ApJ SS, 63, 661


\bibitem[Neuforge-Verheecke et al.(2001)]{guzik1} Neuforge-Verheecke, C., Guzik, J. A., Keady, J. J., Magee, N. H., Bradley, P. A., \& Noels, A. 2001, ApJ, 561, 450

\bibitem[Neuforge et al.(2001)]{guzik2} Neuforge, C., Goriely, S. Guzik, J. A., Bradley, P. A., \& Swenson, F. J. 2001, ApJ, 550, 493

{\color{red}
\bibitem[Pain et al.(2015)]{scorcg} Pain, J.-C., \& Gilleron, F. 2015, High Energy Density Phys., 15, 30 }
%

\bibitem[Paxton et~al.(2015)]{MESA}{Paxton, B. et al.,} 2015, ApJ, 220, 15

\bibitem[Perry et al.(1991)]{perry} Perry, T. S., Davidson, S. J., Serduke, F. J.,  Bach, D. R., Smith, C. C., Foster, J.~M., Doyas, R.~J., 
Ward, R.~A., Iglesias, C. A., Rogers, F. J., Abdallah, Jr., J., Stewart, R.~E., Kilkenny, J.~D., \& Lee, R.~W. 1991, Phys. Rev. Letts., 67, 3784


\bibitem[Potekhin \& Chabrier(2000)]{pc00} Potekhin, A. Y., \& Chabrier, G. 2000, Phys. Rev. E,  62, 8554


\bibitem[Potekhin et al.(2002)]{pcg02} Potekhin, A. Y., Chabrier, G., \& Gilles, D. 2002, Phys. Rev. E, 65, 036412

\bibitem[Rogers \& Iglesias(1992)]{opal1} Rogers, F. J., \& Iglesias, C. A. 1992, ApJS, 79, 507

\bibitem[Sampson(1959)]{samp} Sampson, D. H. 1959, ApJ, 129, 734

\bibitem[Schwerdtfeger(2006)]{nb} Schwerdtfeger, X. 2006, Atomic Static Dipole Polarizabilities, in Computational
Aspects of Electric Polarizability Calculations, (IOS Press) \& {\it http://ctcp.massey.ac.nz/Tablepol-2.8.pdf}

\bibitem[Seaton et al.(1994)]{op0} Seaton, M. J., Yan, Y., Mihalas, D., \& Pradhan, A. K. 1994, MNRAS, 266, 805

\bibitem[Seaton \& Badnell(2004)]{op1}  Seaton, M. J., \& Badnell, N. R. 2004, MNRAS, 354, 457

\bibitem[Serenelli et al.(2009)]{serenelli09} Serenelli, A. M., Basu, S., Ferguson, J. W., \& Asplund, M. 2009, ApJ, 705, L123

\bibitem[Somerville(1965)]{somer} Somerville, W. B. 1965, ApJ, 141, 811

\bibitem[Springer et al.(1992)]{springer} Springer, P. T., Fields, D. J., Wilson, B. G., Nash, J. K., Goldstein, W. H., Iglesias,~C.~A., Rogers,~F.~J.,
Swenson,~J.~K., Chen, M. H., Bar-Shalom, A., \& Stewart,~R.~E. 1992, Phys. Rev. Letts., 69, 3735

\bibitem[Tarafdar \& Vardya(1969)]{TV69} Tarafdar, S. P., \& Vardya, M. S. 1969, MNRAS, 145, 171

\bibitem[Turck-Chi\`eze et al.(2013)]{stc13}   Turck-Chi\`eze, S., Gilles, D., Le Pennec, M., Blenski, T., Thais, F., Bastiani-Ceccotti, S., Blancard, C.,
Busquet, M., Caillaud, T., Colgan, J., Coss\`e, P., Delahaye, F., Ducret, J. E., Faussurier, G., Fontes, C. J.,  Gilleron, F.,
Guzik, J., Harris, J. W., Kilcrease, D. P., Loisel, G., Magee, N. H., Pain, J. C., Reverdin, C., Silvert, V., Villette, B., \&
Zeippen, C. J. 2013,
High Energy Density Phys., 9, 473

\bibitem[Walczak et al.(2015)]{walczak15} Walczak, P., Fontes, C. J., Colgan, J., Kilcrease, D. P., \& Guzik, J. A. 2015, A\&A, 580, L9

\bibitem[Weiss et al.(2004)]{cg04} Weiss, A., Hillebrandt, W., Thomas, H.-C., \& Ritter, H. 2004, Cox \& Giuli's principles of stellar
structure (Extended Second Edition, Cambridge Scientific Publishing)

\bibitem[Winhart et al.(1996)]{winhart} Winhart, G., Eidmann, K., Iglesias, C. A., \& Bar-Shalom, A. 1996, Phys. Rev. E, 53, R1332

\bibitem[Zeiss et al.(1977)]{Zeiss77} Zeiss, G. D. \& Meath, W. J. 1977, Mol. Phys., 33, 1155

\end{thebibliography}
\end{document}